\documentclass[journal, twocolumn]{IEEEtran}
\usepackage{cite}
\usepackage{graphicx}
\usepackage[cmex10]{amsmath}
\usepackage{color,soul}
\usepackage{setspace}
\usepackage{amsthm,amssymb}
\usepackage{mathrsfs} 
\usepackage{algorithm,algorithmic}
\newcounter{mytempeqncnt}
\begin{document}
\title{SPAD-Based Optical Wireless Communication with Signal Pre-Distortion and Noise Normalization}
\author{Shenjie Huang and Majid Safari
	\thanks{The authors are with the School of Engineering, the University of Edinburgh, Edinburgh EH9 3JL, U.K. (e-mail: \{shenjie.huang, Majid.safari\}@ed.ac.uk). This work was supported by Engineering and Physical Sciences Research Council (EPSRC) under Grant EP/R023123/1 (ARROW) and EP/S016570/1 (TOWS). }
}
\maketitle
\begin{abstract}
	In recent years, there has been a growing interest in exploring the application of single-photon avalanche diode (SPAD) in optical wireless communication (OWC). 
	As a photon counting detector, SPAD can provide much higher sensitivity compared to the other commonly used photodetectors. However, SPAD-based receivers suffer from significant dead-time-induced non-linear distortion and signal dependent noise. In this work, we propose a novel SPAD-based OWC system in which the non-linear distortion caused by dead time can be successfully eliminated by the pre-distortion of the signal at the transmitter. In addition, another system with joint pre-distortion and noise normalization functionality is  proposed. Thanks to the additional noise normalization process, for the transformed signal at the receiver, the originally signal dependent noise becomes signal independent so that the conventional signal detection techniques designed for AWGN channels can be employed to decode the signal. {Our numerical results demonstrate the superiority of the proposed SPAD-based systems compared to the existing systems in terms of BER performance and achievable data rate.} 
\end{abstract}
\begin{IEEEkeywords}
Noise normalization, optical wireless communication, single photon avalanche diode.
\end{IEEEkeywords}
\section{Introduction}
In recent years, there is an increase of interest in employing the single-photon avalanche
diode (SPAD) in optical wireless communications (OWC) \cite{Chitnis:14,Khalighi17,Huang20,HuangTime}. To achieve a SPAD receiver, the conventional photodiode should be biased above the breakdown voltage so that it operates in the so-called Geiger mode. One of the main advantages of the SPAD lies in its high sensitivity compared to the other commonly used photodetectors such as PIN photodiode and avalanche photodiode (APD). As summarized in  \cite{Zimmermann}, the SPAD-based receivers can achieve sensitivity distance to the quantum limit down to only $12$ dB; however, for APD-based receivers the corresponding distance is around $20$ dB. In the literature, the application of SPAD receivers in visible light communication (VLC) has been widely investigated \cite{Chitnis:14, Fisher, Zhang:18,Ahmed20}. For instance, by using the off-the-shelf SPAD receiver, an achievable data rate of more than $2$ Gbps with BER of $10^{-3}$ is demonstrated in \cite{Ahmed20}. Some other works also investigate the application of SPAD-based receivers in underwater wireless optical communication (UWOC) \cite{Khalighi17,Khalighi192} and free-space optical communication (FSO) \cite{Huang202}.     

Although SPAD-based receiver can provide excellent single photon sensitivity and picosecond temporal resolution, it suffers from some non-ideal effects such as dead time, crosstalk and afterpulsing. In particular, dead time can introduce significant communication performance degradation which needs to be carefully addressed. Dead time refers to the short
time period of several nanoseconds after each triggered avalanche when the SPAD is getting quenched and hence is unable to detect photons \cite{Khalighi192}. Based on different employed quenching circuits,
SPAD is usually classified into two main types, i.e., active quenching (AQ) and passive quenching (PQ) SPAD. {In PQ SPAD, there is no active circuitry and the avalanche current quenches itself simply by flowing in the quenching resistor \cite{Cova:96}. On the other hand, for AQ SPAD, the rise of the avalanche pulse is sensed by the active circuitry which forces the quenching and reset the SPAD after an accurately controlled hold-off time \cite{Chitnis:14}. 
For AQ SPADs the dead time remains constant, but for PQ SPADs, due to the paralysis property, the photon arrivals during the dead time can extend its duration \cite{Cova:96}. Because of the dead time and limited SPAD array size, both AQ and PQ SPAD have non-linear photon transfer curves. The difference is that for the former, with the increase of the received photon rate, the detected photon rate firstly linearly increases and then saturates at a maximal value, but for the latter it firstly increases and then decreases \cite{eisele2011185}. Such photon transfer property of SPAD could result in significant non-linear distortion to the received signal.} To improve the performance of  SPAD-based receivers in the presence of dead time, a novel detection scheme by utilizing not only the photon count information but also the photon arrival information is proposed in \cite{Huang20}. In addition, a SPAD receiver with variable optical attenuator which can adaptively control the incident photon rate of the SPAD array is investigated in \cite{Huang202}. However, all these techniques require additional hardware at the receiver which inevitably increases the complexity of the receiver design. It is hence crucial to propose cost-effective signal processing techniques for SPAD-based systems to realize performance improvement without any additional hardware requirements.   

Some works in the literature have investigated the signal processing techniques to enhance the performance of OWC systems with signal-dependent noise whose power is proportional to the amplitude of the received signal \cite{Safarinoise,Gao:16}. In \cite{Safarinoise}, the signalling in the square-root domain is proposed which can transform the signal-dependent noise into signal-independent one to simplify the decoding process and improve the BER performance. Such transformation has been further applied to SPAD-based OWC systems \cite{Khalighi192} and optical fibre communication \cite{Iman17,Iman172,Chenyu20}. In addition, in \cite{Gao:16} an improved constellation and decoding threshold design for such signal dependent channel is obtained by solving an optimization problem aiming at minimizing the error probabilities. However, due to the presence of dead time, the signal-dependent noise effect of SPAD receivers is much more complicated than the one described above \cite{elhamJ}. As a result, employing the aforementioned techniques directly in SPAD-based systems cannot achieve the optimal performance. Furthermore, besides the distinct signal-dependent noise, the dead time of SPAD also introduces an additional non-linear signal distortion which makes the channel model different from the simple linear channel model considered in the above works. Therefore, some novel signal processing techniques designed based on the unique properties of SPAD should be proposed to enhance the performance. 

{In this work, two novel SPAD-based OWC systems are proposed to address the dead time effects using signal processing techniques, which, to the best of authors' knowledge, have not been investigated before.  The systems are proposed for general SPAD  OWC links and can be applied to various links such as FSO, VLC and UWOC. {Designed based on the unique characteristics of the SPAD, the proposed systems can achieve superior performance over existing systems.} The main contributions of this paper are summarized as follows: 
\begin{itemize}
	\item A system in which the non-linear distortion caused by the dead time can be successfully eliminated through the pre-distortion of the signal is proposed. 
	\item A system with joint pre-distortion and noise normalization functionality is further proposed in which not only the SPAD nonlinearity effect but also the signal dependency of the noise can be eliminated. The employed noise normalization process can transform the original signal dependent noise of the SPAD output into signal independent one so that the conventional signal detection techniques designed for AWGN channels can be easily employed to decode the received signal. Note that the signal dependent noise of SPAD-based systems cannot be normalized by the prior noise normalization techniques due to dead time effects \cite{Safarinoise,Khalighi192}.
	\item The optimal transmitted multilevel signals of the proposed systems which minimize the BER performance subject to the considered constraints, i.e., the non-negative transmitted signal constraint, the average transmitted power constraint and the peak received power constraint, are derived. 
	\item {Extensive numerical results demonstrate that the proposed systems can achieve superior BER and achievable data rate performance compared to the existing systems, i.e., system with uniform signalling \cite{proakis2001digital,Chen21} and system with square-root transform \cite{Safarinoise,Khalighi192}.}   
	\item The application of the proposed systems in FSO links is investigated which presents the effectiveness of the proposed systems in practical OWC systems.
\end{itemize}}

The rest of this paper is organized as follows. Section \ref{channel} introduces channel model of the SPAD-based OWC systems. The performance of the two commonly used systems is discussed in Section \ref{SOTA}. Section \ref{PS} presents the proposed two SPAD-based systems, i.e., system with pre-distortion and system with joint pre-distortion and noise normalization, and the corresponding BER expressions are derived. Later, the numerical results and discussion are presented in Section  \ref{numer}.  Finally, we conclude this paper in Section \ref{con}. 

\section{Channel Model}\label{channel}
We consider that single-carrier pulse amplitude modulation (PAM) scheme is employed in the considered SPAD-based OWC systems, although the proposed idea in this work can also be extended to systems with other modulation schemes, e.g., OFDM. The transmitted $M$-ary PAM message vector can be expressed as
\begin{equation}\label{2}
\mathbf{s}=\left[0, 1, 2 ,\cdots,(M-1)\right]^T.
\end{equation}
Assuming that after mapping the corresponding vector of the transmitted photon rates is given by
\begin{equation}\label{Is}
I(\mathbf{s})=\left[I(0), \,I(1),\,I(2) ,\cdots,\,I(M-1)\right]^T,
\end{equation}
the average transmitted optical power can be expressed as 
\begin{equation}\label{AveP}
P_T=h\nu E\left[I(\mathbf{s})\right]=\frac{h\nu}{M}\sum_{m=0}^{M-1}I(m),
\end{equation}
where $h\nu$ is the photon energy. Considering that the  loss introduced by the channel is $\alpha$, the average received signal power can be written as 
\begin{equation}
P_R=\alpha P_T=\frac{\alpha h\nu}{M}\sum_{m=0}^{M-1}I(m).
\end{equation}

{The channel loss $\alpha$ depends on the investigated OWC systems. For instance, in FSO systems, the channel loss is random which can be expressed as \cite{Jamali16}
\begin{equation}\label{alphaFSO}
	\alpha_\mathrm{FSO}=h_fh_g=h_f \left[\mathrm{erf}\left(\frac{\sqrt{\pi}\varphi}{2\sqrt{2}\phi L}\right)\right]^2,
\end{equation}		
where $h_g$ denotes the geometric loss induced by diffraction of Gaussian beam, $h_f$ refers to the intensity fluctuation caused by atmospheric turbulence, $\varphi$ is the receiver aperture diameter, $\phi$ is the beam divergence angle at the transmitter, and $L$ refers to the link distance. The intensity fading $h_{f}$ can be modelled by Gamma-Gamma distribution given by \cite{Huang202}
\begin{equation}
	f_{h_{f}}(x)=\frac{2\left(\zeta \beta \right)^{\left(\zeta+\beta\right)/2}}{\Gamma(\zeta)\Gamma(\beta)}{x}^{\left(\zeta+\beta\right)/2-1} K_{\zeta-\beta}\left(2\sqrt{\zeta\beta x}\right),
\end{equation}
where $\Gamma(\cdot)$ is the Gamma function, $K_p(\cdot)$ is the modified Bessel function of the second kind, and the parameter $\zeta$ and $\beta$ are given by
\begin{align}
	\zeta&=\bigg[\mathrm{exp}\bigg( \frac{0.49\chi^2}{\left(1+\!0.18\vartheta^2+0.56\chi^{12/5}\right)^{7/6}}\bigg)-1\bigg]^{-1},\nonumber\\
	\beta&=\bigg[\mathrm{exp}\bigg( \frac{0.51\chi^2\left(1\!+\!0.69\chi^{12/5}\right)^{-5/6}}{\left(1+\!0.9\vartheta^2+0.62\vartheta^2\chi^{12/5}\right)^{5/6}}\bigg)-1\bigg]^{-1},
\end{align}
respectively, with $\chi^2=0.5C_n^2k^{7/6}L^{11/6}$, $\vartheta^2=k\varphi^2/4L$ and $k=2\pi/\lambda_\mathrm{op}$ where $\lambda_\mathrm{op}$ denotes the light wavelength and $C_n^2$ refers to the turbulence refraction structure parameter. On the other hand, for VLC systems with the line-of-sight (LOS) links, the channel loss when the angle of incidence is within the receiver field-of-view (FOV) can be expressed as \cite{Kahn}
\begin{equation}
	\alpha_\mathrm{VLC}=\frac{A_d}{L^2} R_o(\psi_t) G(\psi_r) \mathrm{cos}(\psi_r),
\end{equation}
where $A_d$ is the physical area of the detector, $\psi_t$ refers to the radiance angle,  $R_o(\psi_t)$ is the radiant intensity, $\psi_r$ is the angle of incidence at the receiver, and $G(\psi_r)$ denotes the concentrator gain.  }

In order to achieve higher data rates and additional protection against the background light, SPAD-based receivers in OWC systems are usually implemented in arrays \cite{Khalighi192,Ahmed20,Zhang:18}. {The detected photon counts of individual SPADs in the arrays are superimposed to form a single total photon count output. Considering that SPAD array receivers are very dense, such equal gain combining method is preferable which can greatly simplify the receiver design   \cite{ON,Fisher,Chitnis:14}.} When the $m$th symbol is transmitted at the transmitter, the received photon rate for each SPAD in the array is given by 
\begin{equation}\label{lamm}
\lambda_m=\frac{\Upsilon\left(P_{R,m}+P_b\right)}{Nh\nu}=\frac{\Upsilon\left[\alpha I(m)+\lambda_{b,\mathrm{tot}} \right]}{N},
\end{equation}
{where $\Upsilon$ is the SPAD photon detection efficiency (PDE) which is defined as the product of the fill factor (ratio of the active area to the physical area), avalanche probability and the quantum efficiency.} $P_{R,m}$ denotes the received optical power when the $m$th symbol is transmitted, $P_b$ refers to the total received background power, $N$ denotes the number of SPAD pixels in the array receiver, and $\lambda_{b,\mathrm{tot}}=P_b/h\nu$ is the total received background photon rate. Considering that the employed SPAD receiver is PQ-based, when the received signal is with photon rate of $\lambda_m$, the output photocount distribution of each SPAD pixel can be approximated as sub-Poisson distribution with mean \cite{omote,eisele2011185,Zou19}
\begin{equation}\label{umsingle}
\mu_{m,\mathrm{single}}=\lambda_m T_s e^{-\lambda_m T_d},
\end{equation} 
and variance \cite{omote,daniel2000mean}
\begin{align}\label{var1}
\sigma&_{m,\mathrm{single}}^2=\nonumber\\
&\begin{cases}
\mu_{m,\mathrm{single}}\!-\!\mu_{m,\mathrm{single}}^2, \, & T_s\leq T_d,\\
\mu_{m,\mathrm{single}}\!-\!\mu_{m,\mathrm{single}}^2\left[1-\left(1-\frac{T_d}{T_s}\right)^2\right], \, & T_s> T_d,
\end{cases}
\end{align}
where $T_s$ and $T_d$ represent the symbol duration and dead time, respectively. Since the variance of the photocount is less than its mean, the above distribution shows the sub-Poisson characteristics.  {Note that although AQ SPAD can offer fast reset and well-defined dead time, SPAD arrays with active quenching are with high complexity and cost, hence they are usually designed with small array sizes \cite{Chitnis:14,Ouh20}. On the other hand, PQ SPAD benefits from its simpler circuit design, higher PDE and low cost, therefore the commercial low-cost photon counting receivers with large array sizes are usually with PQ SPADs \cite{ON,Ahmed20}.}  In this work, we hence consider that the SPAD is PQ-based. Equation (\ref{umsingle}) presents that dead time results in the paralysis property of the PQ SPAD and introduces the non-linear distortion to the received signal. In addition, it is demonstrated in (\ref{var1}) that the noise variance of the SPAD output is signal dependent. Note that this signal dependent noise is different from the traditional one investigated in the literature whose power is proportional to the signal amplitude \cite{Safarinoise,Gao:16}. For a SPAD array detector with sufficient large number of pixels, based on the central limit theorem, the total output photocount can be approximated as Gaussian distributed random variable \cite{Khalighi17,elham152} with mean 
\begin{equation}\label{mu}
\mu_{m}=N\mu_{m,\mathrm{single}},
\end{equation} 
and variance
\begin{equation}\label{sigma}
\sigma_{m}^2=N\sigma_{m,\mathrm{single}}^2.
\end{equation}  

To make a fair comparison among the different systems considered in this work, we employ three constraints on the transmitted photon rates (\ref{Is}). Due to the considered intensity modulation direct detection (IM/DD) modulation, the photon rate $I(m)$ should be non-negative. To satisfy this requirement and ensure that the largest dynamic range is utilized, the first constraint is given by 
\begin{equation}\label{1con}
I(0)=0.
\end{equation}
The second constraint is the average power constraint that $P_T\leq P_\mathrm{ave}$ where $P_\mathrm{ave}$ denotes the average power limit of the transmitter. Invoking the expression of $P_T$ given in (\ref{AveP}), this constraint can be expressed as
\begin{equation}\label{con2}
\frac{h\nu}{M}\sum_{m=0}^{M-1}I(m)\leq P_\mathrm{ave}.
\end{equation}
The third considered constraint is the peak power constraint aiming to avoid the strong SPAD non-linear distortion. As illustrated in (\ref{umsingle}), due to the presence of dead time, with the increase of the received photon rate the detected average photon count firstly increases and then decreases. The received photon rate which gives the highest detected photon count, also known as the saturation point, is given by $1/T_d$ \cite{Huang202}. A peak received power constraint can be put to limit the received photon rate below this saturation point, i.e., 
\begin{equation}\label{lamM1}
\lambda_{M-1}\leq \frac{1}{T_d}. 
\end{equation}
Invoking the relationship between the transmitted and received photon rates given in (\ref{lamm}), (\ref{lamM1}) can be rewritten as the constraint on the peak transmitted photon rate 
\begin{equation}\label{PeakP}
I\left(M-1\right)\leq \frac{N}{\alpha T_d\Upsilon }-\frac{\lambda_{b,\mathrm{tot}}}{\alpha}.
\end{equation}

\begin{figure*}[!t]
	\normalsize
	\setcounter{mytempeqncnt}{\value{equation}}
	\setcounter{equation}{23}
	\begin{equation}\label{thre}
		\delta_m=\frac{\frac{\mu_m}{\sigma_m^2}-\frac{\mu_{m+1}}{\sigma_{m+1}^2}+\sqrt{\left(\frac{\mu_m}{\sigma_m^2}-\frac{\mu_{m+1}}{\sigma_{m+1}^2}\right)^2-\left(\frac{1}{\sigma_m^2}-\frac{1}{\sigma_{m+1}^2}\right)\left\{\left(\frac{\mu_m^2}{\sigma_m^2}-\frac{\mu_{m+1}^2}{\sigma_{m+1}^2}\right)+2\,\mathrm{ln}\left(\frac{\sigma_m}{\sigma_{m+1}}\right)\right\}}}{\frac{1}{\sigma_m^2}-\frac{1}{\sigma_{m+1}^2}}. 
	\end{equation}
	\setcounter{equation}{\value{mytempeqncnt}}
	\hrulefill	
\end{figure*}

\section{{SPAD-Based OWC Systems}} \label{SOTA}
\subsection{System with Uniform Signalling}\label{Uniform}
The system with uniform signalling is the simplest and the most common one \cite{Safarinoise,Chen21,proakis2001digital}. In such system, the transmitted photon rates are uniformly spaced which can be expressed as
\begin{equation}\label{Iuni}
I_\mathrm{uni}(\mathbf{s})=\left[0, d, 2d ,\cdots,(M-1)d\,\right]^T,
\end{equation}
where $d$ refers to the constellation level separation at the transmitter. To make sure that the average transmitted power constraint (\ref{con2}) is satisfied, the limitation of $d$ is can be expressed as 
\begin{equation}
d\leq \frac{2P_\mathrm{ave}}{h\nu (M-1)}.
\end{equation}
In addition, the peak power constraint given in (\ref{PeakP}) can be rewritten as
\begin{equation}
d\leq \frac{1}{M-1}\left[\frac{N}{\alpha T_d\Upsilon }-\frac{\lambda_{b,\mathrm{tot}}}{\alpha}\right].
\end{equation}
{To ensure that the above constraints are all satisfied, the maximum constellation level separation at the transmitter is given by}
\begin{equation}
d_{\mathrm{uni}}^*\!=\!\mathrm{min}\left\{\frac{2P_\mathrm{ave}}{h\nu (M-1)},\frac{1}{M-1}\left[\frac{N}{\alpha T_d\Upsilon }\!-\!\frac{\lambda_{b,\mathrm{tot}}}{\alpha}\right]\right\}.
\end{equation}
{For normal OWC systems, larger $d$ is preferable which can result in better BER performance \cite{Safarinoise,Kahn,Zhang:18}, thus the selected constellation level distance is given by $d_{\mathrm{uni}}^*$.} At the receiver, the corresponding received photon rate for each SPAD pixel when the $m$th symbol is transmitted can be written as  
\begin{equation}\label{lamuni}
\lambda_{\mathrm{uni},m}=\frac{\Upsilon\left[\alpha md_{\mathrm{uni}}^*+\lambda_{b,\mathrm{tot}} \right]}{N}, \,\, \mathrm{with} \quad m \in [0, M-1].
\end{equation}

By substituting (\ref{lamuni}) into (\ref{mu}) and (\ref{sigma}), the mean and variance of the SPAD output signal can be achieved. 
Since in the considered system the received constellation levels are not equally space and the noise variance is signal dependent, the commonly used BER expression for PAM-based AWGN channel cannot be employed. Based on the maximum likelihood (ML) rule, the accurate BER for such system can be expressed as \cite{Safarinoise} 
\begin{equation}\label{Pebuniform}
P_\mathrm{eb}\!=\! \frac{1}{M\mathrm{log}_2(M)}\sum_{m=0}^{M-2}Q\left(\frac{\delta_m-\mu_m}{\sigma_m}\right)\!+\!Q\left(\frac{\mu_{m+1}\!-\!\delta_m}{\sigma_{m+1}}\right),
\end{equation}
\stepcounter{equation}
where $\delta_m$ denotes the optimal thresholds calculated based on the moments $\mu_m$ and $\sigma_m^2$ which are given by (\ref{thre}).

\subsection{System with Square-Root Transform}\label{SQRT}
The square-root transform (SQRT) has been applied to the general OWC systems \cite{Safarinoise,Gao:16,Khalighi192} and fibre optical communication systems \cite{Iman17,Chenyu20}. In this section, we will briefly introduce the SPAD-based system with SQRT.

For system with SQRT, a square-root operation is applied to the received signal and at the transmitter an inverse transform, i.e., square operation,  is employed. Therefore, the transmitted photon rate vector can be expressed as
\begin{equation}\label{ImSQRT}
I_\mathrm{sqrt}(\mathbf{s})=\left[0, d, 4d ,\cdots,(M-1)^2d\right]^T.
\end{equation}
Note that different from the photon rates given in (\ref{Iuni}), here the transmitted photon rates are not equally spaced. Now let's consider the optimal selection of $d$ in this system. With the considered photon rate transmission, the average transmitted power constraint (\ref{con2}) can be rewritten as
\begin{equation}
d\leq\frac{6P_\mathrm{ave}}{h\nu (M-1)(2M-1)}. 
\end{equation}
Furthermore, the peak power constraint introduced by the SPAD receiver nonlinearity (\ref{PeakP}) results in 
\begin{equation}
d\leq \frac{1}{(M-1)^2}\left[\frac{N}{\alpha T_d\Upsilon }-\frac{\lambda_{b,\mathrm{tot}}}{\alpha}\right].
\end{equation}
{Therefore, similar to the system with uniform signalling, the selected value of $d$ at the transmitter which satisfies the above constraints can be expressed as}
\begin{align}
d&_{\mathrm{sqrt}}^*=\\
&\mathrm{min}\!\left\{\frac{6P_\mathrm{ave}}{h\nu (M\!-\!1)(2M\!-\!1)},\frac{1}{(M\!-\!1)^2}\left[\frac{N}{\alpha T_d\Upsilon }-\frac{\lambda_{b,\mathrm{tot}}}{\alpha}\right]\right\}\nonumber.
\end{align}
By substituting $d_{\mathrm{sqrt}}^*$ into (\ref{ImSQRT}), the designed transmitted photon rate can be determined. Invoking (\ref{lamm}) the corresponding received photon rate for each SPAD can be expressed as 
\begin{equation}
\lambda_{\mathrm{sqrt},m}=\frac{\Upsilon\left[\alpha m^2d_{\mathrm{sqrt}}^*+\lambda_{b,\mathrm{tot}} \right]}{N}.
\end{equation}
After taking the square-root transform at the receiver, the mean of the transformed signal can be approximated as 
\begin{equation}\label{umtil}
\mu_{\mathrm{sqrt},m}=\sqrt{N\lambda_{\mathrm{sqrt},m}T_s} e^{-\frac{1}{2}\lambda_{\mathrm{sqrt},m} T_d}.
\end{equation}
When the noise variance has linear relationship with the signal amplitude, the system with SQRT can achieve the equidistant transformed received signal with normalized noise variance \cite{Safarinoise}. However, since the relationship between the noise variance and signal amplitude of the SPAD output (as shown in (\ref{var1})) is not linear, SPAD-based system with SQRT cannot achieve the optimal performance. As a result, as illustrated in (\ref{umtil}) the transformed PAM signal is not equidistant. In addition, the noise variance of the transformed signal still keeps its signal dependency. Therefore, (\ref{Pebuniform}) should again be employed to evaluate the BER performance of the considered system. Note that although the mean of the transformed signal is given by (\ref{umtil}), the expression of its variance is analytically intractable. In order to achieve the decoding threshold (\ref{thre}), the variance of the transformed signal should be calculated numerically through the simulation.

\section{The Proposed Systems}\label{PS}
{The prior systems introduced in Section \ref{SOTA} are designed without considering the unique characteristics of the SPAD receivers, thus their performance is strongly degraded by the dead time effects.} In this section, we propose two novel systems aiming to mitigate the dead time effects and improve the communication performance. 
\subsection{System with Pre-Distortion}\label{sysPreD}
	\begin{figure}[!t]
	\centering\includegraphics[width=0.46\textwidth]{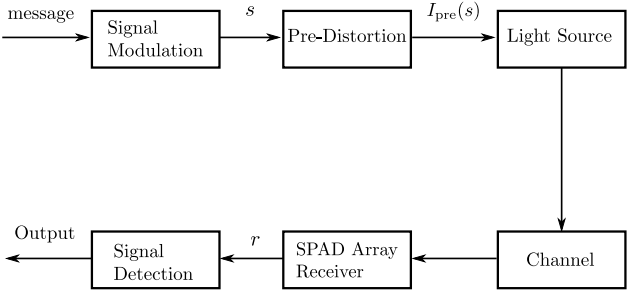}
	\caption{The schematic of the proposed SPAD-based OWC system with signal pre-distortion.} 
	\label{preD}
\end{figure}
The schematic of the first proposed SPAD-based system is shown in Fig. \ref{preD}. At the transmitter a pre-distortion process is applied to the generated PAM signal  and the output signal is then fed to  the light source. After propagating through the free-space channel, the optical signal is received by the SPAD array at the receiver. {The employed pre-distortion process is designed to compensate the non-linear distortion of the SPAD output induced by the dead time and realize the equidistant received signal.} Note that similar signal pre-distortion idea has been proposed in VLC systems to compensate the nonlinearity effects introduced by LEDs \cite{elgala2009non,Deng18}. However, to the best of the authors' knowledge, its application in SPAD-based systems to combat the dead time effects has not been investigated before. In this section, we will fill this research gap.

Invoking (\ref{lamm}) and (\ref{mu}), the relationship between the transmitted photon rate $I(m)$ and average received photon count of the SPAD array $\mu_m$ can be expressed as  
\begin{equation}\label{mum}
\mu_{m}={\Upsilon\left[\alpha I(m)+\lambda_{b,\mathrm{tot}} \right]} T_s e^{-\frac{\Upsilon\left[\alpha I(m)+\lambda_{b,\mathrm{tot}} \right]T_d}{N} },
\end{equation}
To ensure that the distortion induced by SPAD receiver can be compensated, $I(m)$ in the proposed system, denoted as $I_\mathrm{pre}(m)$, should be designed so that the received PAM signals are equidistant, i.e.,
\begin{equation}\label{preDmum}
\mu_{\mathrm{pre},m}=md_{\mathrm{pre}}+\xi_{\mathrm{pre}},
\end{equation}
where $d_{\mathrm{pre}}$ denotes the PAM level separation of the SPAD output and $\xi_{\mathrm{pre}}$ is a constant value. By substituting (\ref{preDmum}) into (\ref{mum}), the desired transmitted photon rate is given by 
\begin{equation}\label{PreDist}
I_\mathrm{pre}(m)=\frac{-N\mathscr{W}_0\left(-\frac{(md_{\mathrm{pre}}\,+\,\xi_{\mathrm{pre}})\,\,T_d}{NT_s}\right)}{\alpha \Upsilon T_d}-\frac{\lambda_{b,\mathrm{tot}}}{\alpha },
\end{equation}
where $\mathscr{W}_0(x)$ denotes the principal branch of the Lambert $W$ function. {Lambert $W$ function is a multivalued function defined to be the inverse of the function $x=w\,e^{w}$ and  $\mathscr{W}_0(x)$ refers to the branch of the function with value above $-1$ \cite{Blondeau}. } The transmitted photon rate (\ref{PreDist}), or equivalently the pre-distortion transform, should satisfy the constraints (\ref{1con}), (\ref{con2}) and (\ref{PeakP}) mentioned in Section \ref{channel}. 

By substituting (\ref{PreDist}) into (\ref{1con}) and after some mathematical manipulations, the constant $\xi_{\mathrm{pre}}$ can be expressed as
\begin{equation}\label{xipred}
\xi_{\mathrm{pre}}={\lambda_{b,\mathrm{tot}}\Upsilon T_s}e^{-\frac{\lambda_{b,\mathrm{tot}}\Upsilon T_d}{N}}. 
\end{equation}
In addition, the peak power constraint (\ref{PeakP}) can be rewritten as
\begin{equation}
{d_{\mathrm{pre}}}\leq \frac{{\frac{NT_s}{eT_d}}-\xi_{\mathrm{pre}}}{M-1}=d_\mathrm{pre,max}.
\end{equation}
Substituting (\ref{PreDist}) into (\ref{con2}), the average power constraint can be expressed as
\begin{equation}\label{noneq}
\underbrace{\frac{1}{M}\!\!\sum_{m=0}^{M-1}\!\!\mathscr{W}_0\!\left(\!\!-\frac{(md_{\mathrm{pre}}\!+\!\xi_{\mathrm{pre}})T_d}{NT_s}\right)\!}_{\mathscr{L}(d_{\mathrm{pre}})}\!\geq\!\frac{-\Upsilon T_d}{Nh\nu}\!\left(\alpha P_\mathrm{ave}\!+\!\lambda_{b,\mathrm{tot}}h\nu \right),
\end{equation}
where $\mathscr{L}(d_{\mathrm{pre}})$ refers to the function at the left hand side of (\ref{noneq}). As $\mathscr{W}_0(x)$ is a monotonically increasing function for $x\in[-1/e,0]$, $\mathscr{L}(d_{\mathrm{pre}})$ is a monotonically decreasing function with respect to $d_{\mathrm{pre}}$ with a maximal value
\begin{equation}
\mathscr{L}(0)\!=\!\mathscr{W}_0\left(\frac{-\lambda_{b,\mathrm{tot}}\Upsilon T_d}{N}e^{\frac{-\lambda_{b,\mathrm{tot}}\Upsilon T_d}{N}}\right)\!=\!\frac{-\lambda_{b,\mathrm{tot}}\Upsilon T_d}{N},
\end{equation}
and a minimal value $\mathscr{L}(d_\mathrm{pre,max})$. Therefore, the inequality (\ref{noneq}) can be expressed as $d_{\mathrm{pre}}\leq d_{\mathrm{pre},\, \mathrm{root}}$ when $\mathscr{L}\left(d_\mathrm{pre,max}\right)\leq \frac{-\Upsilon T_d}{Nh\nu}\left(\alpha P_\mathrm{ave}+\lambda_{b,\mathrm{tot}}h\nu \right)$ is satisfied and $d_{\mathrm{pre}}\leq d_{\mathrm{pre},\, \mathrm{max}}$ otherwise,
where $d_{\mathrm{pre},\, \mathrm{root}}$ denotes the single positive root of (\ref{noneq}) when the equality holds which can be calculated numerically through the bisection method. {As larger $d_{\mathrm{pre}}$ indicates better BER performance (as proved in Appendix), the optimal $d_{\mathrm{pre}}$, denoted as $d_{\mathrm{pre}}^*$, is achieved at the boundary of the feasible set and hence is expressed as}
\begin{equation}\label{dpred}
d_{\mathrm{pre}}^*\!=\!
\begin{cases}
d_{\mathrm{pre},\, \mathrm{root}}, \,\, \mathscr{L}\left(d_\mathrm{pre,max}\right)\!\leq\! \frac{-\Upsilon T_d}{Nh\nu}\left(\alpha P_\mathrm{ave}+\lambda_{b,\mathrm{tot}}h\nu \right),\\
d_{\mathrm{pre},\, \mathrm{max}}, \,\, \mathrm{otherwise}. 
\end{cases}
\end{equation} 
By substituting the derived $\xi_{\mathrm{pre}}$ in (\ref{xipred}) and $d_{\mathrm{pre}}^*$ in (\ref{dpred}) into (\ref{PreDist}), the pre-distortion transform at the transmitter can be determined.

As shown in (\ref{preDmum}), due to the utilisation of the signal pre-distortion, the constellation levels at the receiver are equidistant with a separation of $d_{\mathrm{pre}}^*$.  Substituting (\ref{preDmum}) into (\ref{var1}) and (\ref{sigma}) the variance of the received signal, denoted as $\sigma_{\mathrm{pre},m}^2$, can be determined which is, however, still signal dependent. Therefore, the BER of the considered system is again given by (\ref{Pebuniform}) with the optimal decoding thresholds calculated based on the derived moments $\mu_{\mathrm{pre},m}$ and $\sigma_{\mathrm{pre},m}^2$. 

\subsection{System with Joint Pre-Distortion and Noise Normalization}\label{proposed}
	\begin{figure}[!t]
	\centering\includegraphics[width=0.47\textwidth]{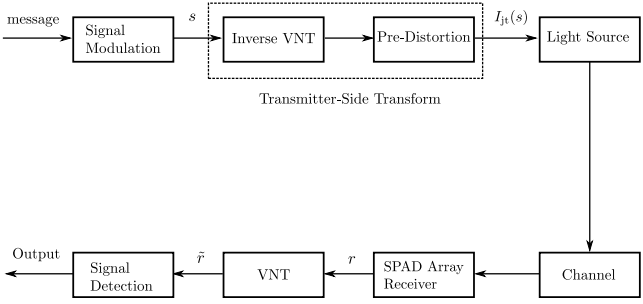}
	\caption{The schematic of the proposed SPAD-based OWC system with joint pre-distortion and noise normalization.} 
	\label{4PAM}
\end{figure}
As illustrated in Section \ref{SQRT}, the system with SQRT cannot successfully normalize the noise variance of SPAD-based OWC systems due to the non-linear relationship between the noise variance and signal amplitude.  In this section, we aim to design a novel SPAD-based OWC system with joint pre-distortion and noise normalization functionality. The schematic of the proposed system is demonstrated in Fig. \ref{4PAM}. Different from the system discussed in Section \ref{sysPreD}, in this system an additional variance normalizing transform (VNT) and inverse VNT are employed at the receiver and transmitter, respectively. The proposed system can not only achieve the equidistant PAM signals at the receiver as the system considered in Section \ref{sysPreD},  but also successfully normalize the noise variance so that the noise becomes signal independent. Thanks to the joint functionality, it is expected that the proposed system outperforms the system with only pre-distortion; however, the latter benefits from its simplicity as it does not require any additional signal processing at the receiver. In the following discussion, the design of the transformations involved in the proposed system will be investigated.

The variance of the output of SPAD pixel given in (\ref{var1}) can be further simplified as 
\begin{equation}\label{sigmasingle}
\sigma_{m,\mathrm{single}}^2=\mu_{m,\mathrm{single}}-\mu_{m,\mathrm{single}}^2\vartheta, 
\end{equation}
where 
\begin{equation}
\vartheta=
1-\left[\left(1-\frac{T_d}{T_s}\right)^+\right]^2,
\end{equation}
with $(x)^+=\mathrm{max}\{x,0\}$ and $\vartheta\in (0,1]$. Denoting the output signal of SPAD array receiver as $r$, as mentioned in  Section \ref{channel} it is approximately Gaussian distributed. Invoking (\ref{sigma}) the variance of $r$ can be expressed as
\begin{equation}\label{sigmam}
\sigma_{m}^2=\mu_{m}-\frac{\vartheta}{N}\mu_{m}^2.
\end{equation} 
Let's firstly consider the design of VNT at the receiver. The VNT is applied to random variables with certain relationships between the variance and mean to generate approximately Gaussian random variables with variance independent of the mean \cite{Safarinoise, Iman17,Prucnal:81,Chenyu20}\footnote{Note that the convergence to Gaussian random variables is shown for the random variables with original gamma distribution \cite{Prucnal:81}, Gaussian distribution \cite{Safarinoise}, noncentral chi-squared distribution \cite{Chenyu20}, and an unknown distribution \cite{Iman17}.}.  Defining the relationship between the mean and variance of the received SPAD signal given in (\ref{sigmam}) as $\sigma_m^2=f^2(\mu_m)$, the function $f(x)$ can be expressed as 
\begin{equation}
f(x)=\sqrt{x-\frac{\vartheta}{N}x^2}.
\end{equation}
The VNT at the receiver aiming to normalize the signal-dependent noise can be expressed as \cite{Safarinoise}
\begin{equation}\label{Tx}
\mathscr{T}(x)=\int \frac{1}{f(x)} \,\mathrm{d}x=\int \frac{1}{\sqrt{x-\frac{\vartheta}{N}x^2}}\,\,\mathrm{d}x.
\end{equation} 
The above integral can be solved analytically as
\begin{equation}\label{Tx2}
\mathscr{T}(x)=-\sqrt{\frac{N}{{\vartheta}}}\,\mathrm{arcsin}\left(-\frac{2\vartheta}{N} x+1\right).
\end{equation}
Denoting the transformed signal after VNT as $\tilde{r}$, i.e., $\tilde{r}=\mathscr{T}(r)$,  it is with mean   $\mu_{\tilde{r}}\simeq\mathscr{T}(\mu_m)$ and variance
$\sigma_{\tilde{r}}^2\simeq 1$ when $\mu_m$ is relatively high. Note that the noise variance of $\tilde{r}$ is signal independent.  As the received signal $r$ is approximately Gaussian distributed, it is possible (with very low probability) that when applying the transformation $\mathscr{T}(r)$, the value $-\frac{2\vartheta}{N} r+1$ is out of the domain of arcsine function, i.e., $[-1,1]$. To address this issue, before applying the transformation, we truncate $r$ such that $r=N/\vartheta$ and $r=0$ when $r>N/\vartheta$ and $r<0$ are satisfied, respectively. 

Now let's turn to the design of the transformation at the transmitter. The PAM signal is transformed into the transmitted photon rate $I_\mathrm{jt}(m)$ and then propagated through the free-space channel. The objective is that for the received signal after VNT the noise variance is signal independent and the constellation levels are equidistant, i.e., 
\begin{equation}\label{obj}
\tilde{r}=\mu_{\tilde{r}}+n_{\tilde{r}}, 
\end{equation}
with
\begin{equation}\label{mutild}
\mu_{\tilde{r}}=md_\mathrm{jt}+\xi_\mathrm{jt},
\end{equation}
where $n_{\tilde{r}}$ is the approximately Gaussian distributed noise with zero mean and normalized variance equal to one, $d_\mathrm{jt}$ refers to the PAM level separation after VNT, and $\xi_\mathrm{jt}$ is a constant value.
To achieve this goal, the transmitter side transformation should inverse both VNT and the non-linear distortion caused by SPAD channel.   Considering that $\mu_{\tilde{r}}\simeq\mathscr{T}(\mu_m)$, (\ref{mutild}) can be rewritten as 
\begin{equation}\label{Tum2}
\mathscr{T}\left(\mu_{m}\right)=md_\mathrm{jt}+\xi_\mathrm{jt}.
\end{equation}
By substituting (\ref{mum}) into (\ref{Tum2}) and after some mathematical manipulations, one can get
\begin{equation}\label{equ}
\mathscr{R}\left(\!-\frac{\Upsilon T_d\left[\alpha I_\mathrm{jt}(m)\!+\!\lambda_{b,\mathrm{tot}} \right]}{N}\!\right) \!=\!\frac{-\mathscr{T}^{-1}(md_\mathrm{jt}\!+\!\xi_\mathrm{jt})T_d}{NT_s},
\end{equation}
where the function $\mathscr{R}(x)=xe^{x}$ and $\mathscr{T}^{-1}(x)$ denotes the inverse function of $\mathscr{T}(x)$ which can be expressed as 
\begin{equation}\label{Tinv}
\mathscr{T}^{-1}(x)=\frac{-N}{2\vartheta}\left[\mathrm{sin}\left(-\sqrt{\frac{\vartheta}{N}}x\right)-1\right].
\end{equation}
Equation (\ref{equ}) can be rewritten as 
\begin{equation}\label{eq2}
\frac{-\Upsilon T_d\left[\alpha I_\mathrm{jt}(m)+\lambda_{b,\mathrm{tot}}  \right]}{N} \!=\!\mathscr{W}_0\left\{\frac{-\mathscr{T}^{-1}(md_\mathrm{jt}+\xi_\mathrm{jt})T_d}{NT_s}\right\}.
\end{equation}
Based on (\ref{eq2}), the desired transmitted photon rate when the $m$th symbol is transmitted can be expressed as
\begin{equation}\label{111}
	I_\mathrm{jt}(m)=\frac{-N\mathscr{W}_0\left\{-\frac{T_d}{N T_s} \mathscr{T}^{-1}    \left(md_\mathrm{jt}+\xi_\mathrm{jt}\right)          \right\}}{\alpha \Upsilon T_d}-\frac{\lambda_{b,\mathrm{tot}}}{\alpha }.
\end{equation}
This transmitted photon rate, or equivalently the transmitter side transformation, has the similar shape to that of the system with only pre-distortion given in (\ref{PreDist}). The difference is that in this proposed system, the generated PAM signal firstly goes through an additional inverse VNT which is given by $-\frac{T_d}{N T_s} \mathscr{T}^{-1}    \left(md_\mathrm{jt}+\xi_\mathrm{jt}\right)$ after which  the pre-distortion transform is applied. Therefore, the transmitter side transform can be divided into two sub-transforms, i.e., inverse VNT and pre-distortion, as illustrated in Fig. \ref{4PAM}. 

In order to apply the derived transformations at the transmitter, the PAM level separation  $d_\mathrm{jt}$ and the constant $\xi_\mathrm{jt}$ in (\ref{111}) should be determined. 
Substituting (\ref{111}) into the first considered constraint (\ref{1con}) and after some mathematical manipulations, one can get 
\begin{equation}\label{xi}
\xi_\mathrm{jt}=-\sqrt{\frac{N}{\vartheta}}\mathrm{arcsin}\left(\frac{-2\vartheta T_s\lambda_{b,\mathrm{tot}}\Upsilon}{N}e^{-\frac{\lambda_{b,\mathrm{tot}}\Upsilon T_d}{N}}+1\right).
\end{equation}
On other hand, the peak power constraint given in (\ref{PeakP}) can be rewritten as
\begin{equation}\label{drmax}
d_\mathrm{jt}\leq \frac{-\sqrt{\frac{N}{\vartheta}} \mathrm{arcsin}\left(-\frac{2\vartheta T_s}{T_d e}+1\right)- \xi_\mathrm{jt}    }{M-1}=d_{\mathrm{jt, \,max}}.
\end{equation}
It can be easily shown that ${\vartheta T_s}/{T_d \,e}\in (0,\frac{1}{e}]$ when $T_d\geq T_s$ and ${\vartheta T_s}/{T_d \,e}\in (\frac{1}{e},\frac{2}{e})$ when $T_d< T_s$. Therefore, in both cases, the term $-{2\vartheta T_s}/{T_d\, e}+1$ in (\ref{drmax}) is within the domain of the arcsine function for real results. Finally, with the expression of $I_\mathrm{jt}(m)$ given in (\ref{111}), the average power constraint (\ref{con2}) can be expressed as
\begin{align}\label{3}
&\underbrace{\frac{1}{M}\sum_{m=0}^{M-1}\mathscr{W}_0\left\{\frac{T_d}{2\vartheta T_s}\left[\mathrm{sin}\left(-\sqrt{\frac{\vartheta}{N}}\left(md_\mathrm{jt}+\xi_\mathrm{jt}\right)\right)-1\right]\right\}}_{\mathscr{L}'(d_\mathrm{jt})}\nonumber\\
&\quad\quad\geq\frac{-\Upsilon T_d}{Nh\nu}\left(\alpha P_\mathrm{ave}+\lambda_{b,\mathrm{tot}}h\nu \right).
\end{align}
Note that we denote the function at the left hand side of the above inequality as $\mathscr{L}'(d_\mathrm{jt})$. It can be proven that  
$\mathscr{L}'(d_\mathrm{jt})$ is a monotonically decreasing function with respect to $d_\mathrm{jt}$ with a maximal value
\begin{equation}
\mathscr{L}'(0)\!=\!\mathscr{W}_0\left(\frac{T_d}{2\vartheta T_s}\!\left[\mathrm{sin}\!\left(\!\!-\sqrt{\frac{\vartheta}{N}}\xi_\mathrm{jt}\right)\!-\!1\right]\right)=\frac{-\lambda_{b,\mathrm{tot}}\Upsilon T_d}{N}.
\end{equation}
Invoking the constraint (\ref{drmax}), the minimal value of $\mathscr{L}'(d_\mathrm{jt})$ is $\mathscr{L}'(d_{\mathrm{jt,\,max}})$. 
{Using the same method employed to derive (\ref{dpred}), the maximal $d_\mathrm{jt}$ subject to the considered constraints, denoted as $d_\mathrm{jt}^*$, can be expressed as}
\begin{equation}\label{dr*}
d_\mathrm{jt}^*=\begin{cases}
		d_{\mathrm{jt},\,\mathrm{root}},  \,\, \mathscr{L}'\left(d_{\mathrm{jt, \,max}}\right)\leq \frac{-\Upsilon T_d}{Nh\nu}\left(\alpha P_\mathrm{ave}+\lambda_{b,\mathrm{tot}}h\nu \right),\\
		d_{\mathrm{jt, \,max}},  \,\, \mathrm{otherwise},
	\end{cases}
\end{equation} 
where $d_{\mathrm{jt},\,\mathrm{root}}$ denotes the single positive root of (\ref{3}) when the equality holds which can be achieved numerically. As explained below, the optimal $d_\mathrm{jt}$ which results in the minimal BER is $d_\mathrm{jt}^*$. Substituting the derived $\xi_\mathrm{jt}$ in (\ref{xi}) and $d_\mathrm{jt}^*$ in (\ref{dr*}) into (\ref{111}), the transformations at the transmitter can be determined.

By utilizing the transformations at transmitter and receiver given by (\ref{111}) and (\ref{Tx2}), respectively, the proposed system has joint pre-distortion and noise normalization functionality.  
Assuming that the noise normalization process is ideal, the transformed signal at the receiver is Gaussian distributed with signal independent noise. {Note that later in Section \ref{numer} the accuracy of this assumption will be verified through numerical simulations.}  In addition, thanks to the signal pre-distortion, constellation levels of the transformed signal are equally spaced. Thus for the proposed system, the channel is approximately AWGN and the conventional signal detection techniques designed for AWGN channel can be applied to recover the original message. {Since for AWGN channel, larger constellation separation $d_\mathrm{jt}$ results in better BER performance \cite{proakis2001digital}, the optimal $d_\mathrm{jt}$ is hence $d_\mathrm{jt}^*$ given in (\ref{dr*}).} The BER of the proposed system can be approximated by the BER for PAM modulation in AWGN channel given by
\begin{equation}\label{Pebprop}
P_{\mathrm{eb},\mathrm{jt}}\!\approx\!\frac{2M-2}{M\mathrm{log}_2(M)}Q\!\left(\frac{d_\mathrm{jt}^*}{2\sigma_{\tilde{r}}}\right)\!=\!\frac{2M-2}{M\mathrm{log}_2(M)}Q\left({\frac{d_\mathrm{jt}^*}{2}}\right).
\end{equation}

{After the VNT at the receiver, the signal distribution is not exactly Gaussian as mentioned before, hence applying  ML decoding based on AWGN approximation results in a sub-optimal BER performance. To achieve the optimal BER performance, one should apply ML decoding to the signal after VNT without employing any approximation. Since the VNT shown in (\ref{Tx2}) is a monotonic function which just performs a one-to-one mapping, the ML decoding of the SPAD output signal is in effect equivalent to that of the signal after VNT without  approximation. 
Because the output of the SPAD is already modelled as Gaussian distributed, applying ML decoding directly to the SPAD output (i.e., signal before VNT) is more mathematically tractable, which will be considered next to find the optimal BER performance.
By substituting the designed transmitted photon rate (\ref{111}) into (\ref{mu}) and (\ref{sigma}), the mean and variance of the SPAD output can be achieved where the variance is signal dependent. As the SPAD output is Gaussian distributed, similar to the BER analysis in Section \ref{SOTA} and Section \ref{sysPreD}, the optimal BER based on ML decoding can be expressed as (\ref{Pebuniform}) with thresholds given by (\ref{thre}). Note that compared to the case of applying ML decoding to the VNT output with AWGN approximation, additional complexity is added to the calculation of the decoding thresholds especially when higher order modulation is considered.	Latter in Section \ref{numer}, these two considered BER performances will be compared.}   

\section{Numerical Results}\label{numer}
\begin{table}
	\renewcommand{\arraystretch}{1.5}
	\caption{The Parameter Setting  \cite{Jamali16,Sarrah}}
	\label{table}
	\centering
	\resizebox{0.44\textwidth}{!}
	{\begin{tabular}{|c|c|c|}
			\hline
			Symbol & Definition & Value\\
			\hline\hline
			$\lambda_\mathrm{op}$ & Optical wavelength & $785$ nm\\
			\hline
			$\Upsilon$ & The PDE of SPAD & $0.18$ \\ 
			\hline 
			$N$&  Number of SPAD pixels in the array & [$2048$, $4096$] \\
			\hline
			$T_d$ & The dead time of SPAD &  $10$ ns\\
			\hline
			\multicolumn{3}{|c|}{FSO link parameters}  \\	
			\hline 
			$\varphi$ & {Receiver lens aperture diameter} & $10$ cm \\
			\hline
			$L$ & Distance between the source and destination & $1500$ m\\
			\hline
			$\phi$ & Laser divergence angle & $2$ mrad \\
			\hline
			$P_{b}$ & Background power & $20$ nW\\ 
			\hline
			$C_n^2$ & Refraction structure index & $[10^{-15},\, 10^{-13}]$ $\mathrm{m}^{-2/3}$ \\
			\hline
			$R$ & Data rate of the FSO link & $1$ Gbps \\
			\hline
	\end{tabular}}
\end{table}
In this section, some numerical results are presented. Unless otherwise mentioned, the parameters used in the simulation are given in Table \ref{table}. {We firstly present the numerical results showing the superiority of the proposed systems in Section \ref{PS} over the existing systems introduced in Section \ref{SOTA}.} Later, the performance improvement by utilizing the proposed systems in a practical FSO link is investigated. 

\subsection{The Superiority of the Proposed Systems}
To show the effectiveness of the propose systems, let's consider a SAPD-based OWC system with a SPAD receiver with $2048$ pixels and a channel loss of $\alpha=-30$ dB. We consider 4-PAM as the modulation scheme, although other PAM modulation schemes can also be employed which can result in similar conclusions. 
\begin{figure}[!t]
	\centering\includegraphics[width=0.5\textwidth]{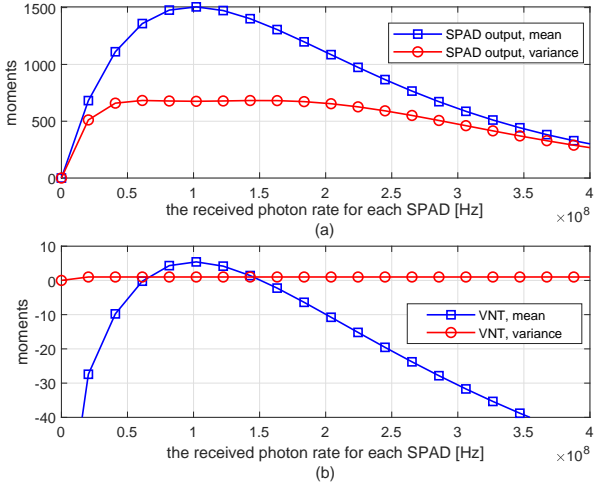}
	\caption{The mean and variance of (a) the original SPAD output signal and (b) the signal after the proposed VNT at the receiver given in (\ref{Tx2}). The symbol duration is $T_s=20$ ns. } 
	\label{noise_norm_Gaussian}
\end{figure}

Firstly, let's consider the effectiveness of the VNT given in (\ref{Tx2}). Fig. \ref{noise_norm_Gaussian} shows the mean and variance of the original SPAD output signal and the signal after the VNT versus the received photon rate. {The moments of the SPAD output signal are given by (\ref{umsingle}) and (\ref{var1}); whereas, the moments of the signal after VNT are achieved numerically through simulation. As shown in Fig. \ref{noise_norm_Gaussian}(a), the noise variance of SPAD output is signal dependent and is less than the mean value which reveals the sub-Poisson characteristics of the SPAD signal.} However, when the VNT is applied, as demonstrated in Fig. \ref{noise_norm_Gaussian}(b), with the increase of received photon rate the noise variance of the transformed signal quickly increases to $1$ and becomes fixed at this value. Therefore, the variance of the noise can be successfully normalized by employing the designed VNT at the receiver.

\begin{figure}[!t]
	\centering\includegraphics[width=0.48\textwidth]{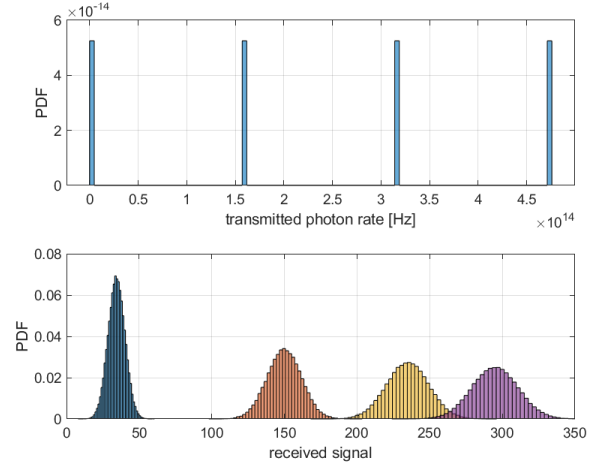}
	\caption{The signal PDFs of the system with uniform signalling when the average transmitted power limit is $P_\mathrm{ave}=60$ $\mu$W, the received background power is $P_b=10$ nW, and symbol duration is $T_s=5$ ns.  } 
	\label{uniform_signalling}
\end{figure}
\begin{figure}[!t]
	\centering\includegraphics[width=0.48\textwidth]{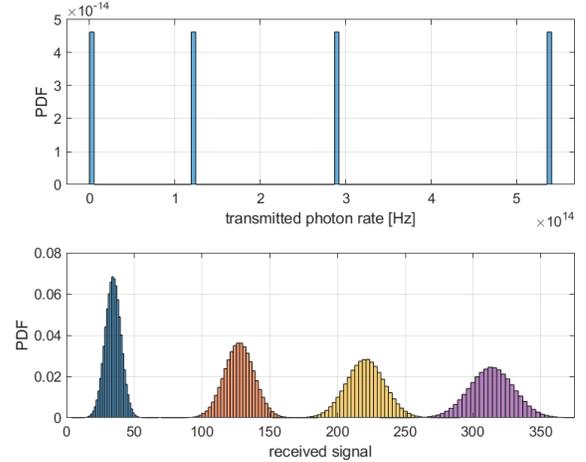}
	\caption{The signal PDFs of the proposed system with signal pre-distortion when the average transmitted power limit is $P_\mathrm{ave}=60$ $\mu$W, the received background power is $P_b=10$ nW, and symbol duration is $T_s=5$ ns.  } 
	\label{PreDis_Gaussian_60uW}
\end{figure}

Figure  \ref{uniform_signalling} presents an example of the PDFs of the transmitted photon rate and the corresponding received signal for the system with uniform signalling. It is shown that even though the transmitted photon rates are uniformly spaced, due to the SPAD-induced non-linear distortion and shot noise effects, the received signal levels are not uniformly spaced and the signal variance is strongly signal dependent. For instance, the distance between the first two received constellation levels is around $115$; however, the distance between the last two levels is only $60$. In addition, the variance of the received signal when the lowest level is transmitted is $34$; whereas, the corresponding variance when the highest level is transmitted increases to $252$.

\begin{figure}[!t]
	\centering\includegraphics[width=0.51\textwidth]{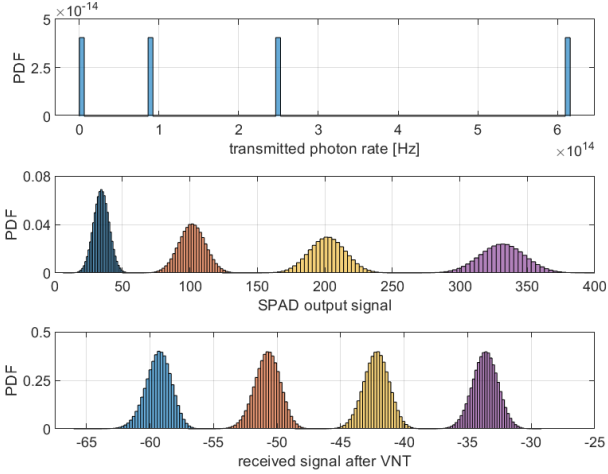}
	\caption{The signal PDFs of the proposed system with joint functionality when the average transmitted power limit is $P_\mathrm{ave}=60$ $\mu$W, the received background power is $P_b=10$ nW,  and symbol duration is $T_s=5$ ns.}
	\label{proposed_signalling}
\end{figure}
Figure  \ref{PreDis_Gaussian_60uW} demonstrates the signal PDFs of the proposed system with pre-distortion. It is shown that with the employed signal pre-distortion at the transmitter, the received constellation levels successfully become equidistant with a separation of $93$. However, the noise is still signal-dependent and higher signal level still results in higher noise variance. On the other hand, the PDFs of the transmitted signal, SPAD output signal and signal after VNT for the proposed system with joint pre-distortion and noise normalization functionality are presented in Fig. \ref{proposed_signalling}. One can observe that, similar to system with only pre-distortion, the constellation levels of the signal after the VNT are also equally spaced (with distance $8.6$). However, the difference is that for such system the noise variances when different signal levels are transmitted are also identical (approximately equal to $1$) hence the noise is signal independent. Therefore, in this considered system both the non-linear distortion and signal-dependent noise of the SPAD output are eliminated, as expected. {It is worth noting that although the output of the SPAD is the detected photon count which is non-negative, the sign of the signal after VNT (\ref{Tx2}) is undetermined which could be negative. However, the negative values of the output signal have no effects on the system performance, because in the decoding process it is the relative distances among the multilevel signals which determine the communication performance rather than the absolute values of the  signals.}

\begin{figure}[!t]
	\centering\includegraphics[width=0.51\textwidth]{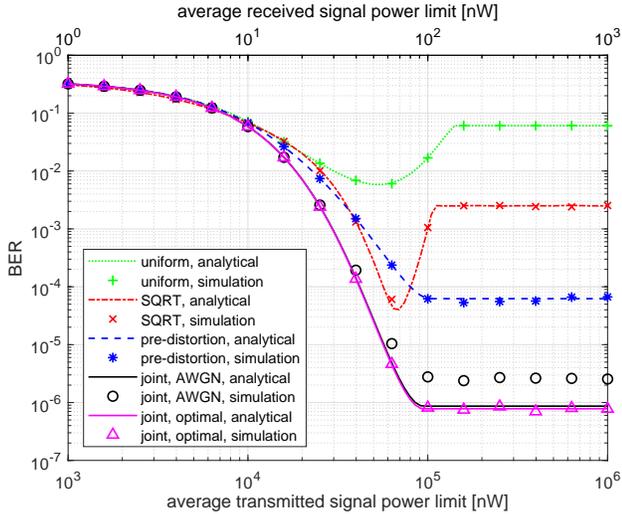}
	\caption{{The BER versus the average transmitted and received optical signal power limits for systems with different schemes when the symbol duration is $T_s=5$ ns and the received background power is $P_b=10$ nW.}} 
	\label{BER_combin_Pb10nW}
\end{figure}
\begin{figure}[!t]
	\centering\includegraphics[width=0.51\textwidth]{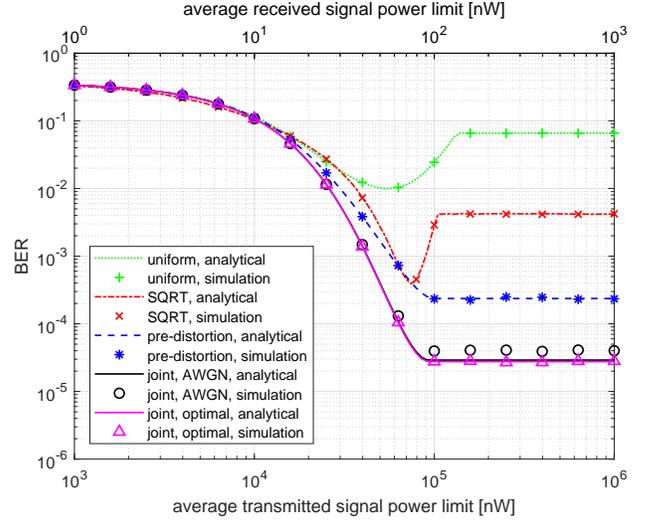}
	\caption{{The BER versus the average transmitted and received optical signal power limits for systems with different schemes when the symbol duration is $T_s=5$ ns and the received background power is $P_b=20$ nW.}}  
	\label{BER_combin_Pb20nW}
\end{figure}
{Figure \ref{BER_combin_Pb10nW} and Figure \ref{BER_combin_Pb20nW} present the BER versus the average transmitted optical power limit $P_\mathrm{ave}$ for the investigated systems under two different background power levels. To give insights into the operational received power of the considered SPAD-based systems, the corresponding average received signal power limit denoted as $\alpha P_\mathrm{ave}$ is also added (see top axis).  Note that these power limits are not always the actual transmitted and received signal powers due to the employed additional peak power constraint given in (\ref{PeakP}).} For the system with uniform signalling, with the increase of the transmitted power limit, the BER firstly decreases and then increases due to the non-linear distortion effects of the SPAD. With further increase of the transmitted power limit, the BER saturates at a fixed value. This is because of the considered peak power constraint  (\ref{PeakP}) which results in the BER performance becomes average power independent in high average power regime. The same BER saturation effect can also be observed in the other systems. The system with SQRT performs slightly better than that with uniform signalling but they are with similar BER shapes. On the other hand, for the proposed systems, i.e., systems with pre-distortion and joint functionality, thanks to the signal pre-distortion which compensates the SPAD non-linear effects at the transmitter, the BER monotonically decreases and then saturates with the increase of the transmitted power. {It is demonstrated that generally these two proposed systems can result in much better BER performance compared to the conventional systems, i.e., system with uniform signalling and SQRT, especially in high power regimes.} {For the system with joint functionality, the performances of both ML decoding in the presence and absence of AWGN approximation are plotted. In the presence of AWGN, the 
BER expression is given by (\ref{Pebprop}); whereas, in the absence of AWGN approximation, the calculation of the optimal BER is based on (\ref{Pebuniform}) according to the discussion at the end of Section \ref{proposed}. As mentioned in Section \ref{proposed}, the advantage of the former is the simpler decoding process. It is shown that the proposed system with joint functionality can achieve the best BER performance among the considered systems over the whole considered optical power regime.
For example, when $P_\mathrm{ave}=100$ $\mu$W and $P_b=10$ nW, the BER values of the systems with uniform signalling, SQRT, and pre-distortion are given by $2\times 10^{-2}$, $10^{-3}$ and $6.2\times 10^{-5}$, respectively. However, the corresponding optimal BER of the proposed system with joint functionality is only $8\times 10^{-7}$. }

{It is also presented in Figure \ref{BER_combin_Pb10nW} and Figure \ref{BER_combin_Pb20nW}  that for the considered systems except the joint functionality system with AWGN approximation, the analytical results exactly match with the simulation ones, which justifies our analytical derivations. For the joint functionality system with AWGN approximation, the analytical result slightly outperforms the simulation result. This is because after the VNT at the receiver, the transformed signal is not exactly Gaussian distributed. As a result, the analytical BER expression given in (\ref{Pebprop}) which is calculated based on ideal AWGN channel cannot perfectly match the corresponding simulation result. However, the small deviation between the analytical and simulation result indicates that this analytical expression is still a good approximation of the simulation result. In addition, it is shown that for the system with joint functionality, the simulated BER with AWGN assumption
is worse than the optimal BER as explained in Section \ref{proposed}, but the gap between them is very small. It is interesting that the optimal BER performance is almost the same as the analytical BER performance assuming AWGN given in (\ref{Pebprop}), which implies that (\ref{Pebprop}) can be a good approximation of the optimal performance. }

\begin{figure}[!t]
	\centering\includegraphics[width=0.5\textwidth]{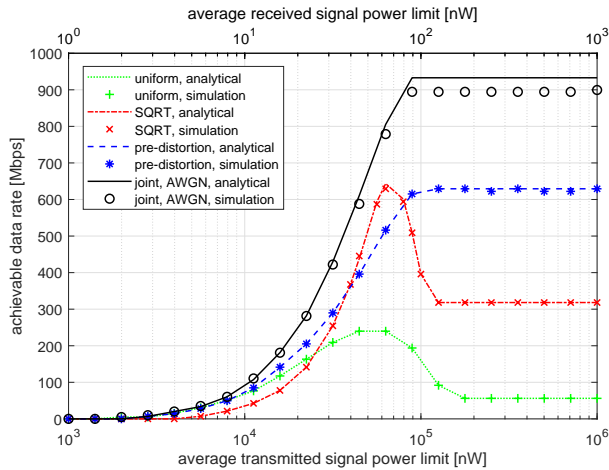}
	\caption{{The achievable data rate versus the average transmitted and received optical signal power limits for different schemes when the BER target is $10^{-3}$ and the received background power is $P_b=10$ nW.}}
	\label{DR_combin_Pb10nW_new}
\end{figure}
By comparing Fig. \ref{BER_combin_Pb10nW} and Fig. \ref{BER_combin_Pb20nW}, it is  presented that the increase of the received background power can result in significant BER degradation to all of the considered systems. In addition, higher background power also leads to the reduction of the performance gap between the system with joint functionality and the system with only pre-distortion. This is because higher background power introduces less available dynamic range and results in the less difference among the variances when different signal levels are transmitted. Consequently, for the system with joint functionality, the performance improvement induced by additional noise normalization reduces, which makes its performance closer to that of the system with only pre-distortion.

The achievable data rates versus the average transmitted and received power limits for  different systems are plotted in Fig. \ref{DR_combin_Pb10nW_new}. {In the simulation, a target BER of $10^{-3}$ is employed which is below the $7\%$ forward error correction (FEC) limits, i.e., $3.8\times 10^{-3}$ \cite{Wang:13,Patel15}.
A vector of the achievable data rates (or equivalently the symbol durations) with relatively small step size is generated. For each given transmitted optical power, the BER performances of the considered systems under different data rate transmission are calculated either analytically or through simulation, and by comparing with the BER target the maximal achievable data rates of various systems can then be determined. Note that for the system with joint functionality, only the performance in the presence of AWGN approximation is presented, as the performance with optimal decoding is similar to that of the system with AWGN as mentioned above.} It is clearly demonstrated that generally the proposed systems can effectively improve the achievable data rate. In particular, the proposed system with joint functionality can achieve the highest data rate over the considered range of optical power. For instance, when the transmitted optical power is $200$ $\mu$W, the achievable data rate of the proposed system with joint functionality is $900$ Mbps; however, the corresponding data rates for system with uniform signalling, SQRT and pre-distortion are only $56$ Mbps, $318$ Mbps and $630$ Mbps, respectively. 

\subsection{The application in Practical FSO Link}
{The proposed signal processing techniques are designed for general SPAD-based OWC systems. However, when applied in different systems, disparate channel effects should be considered, e.g., the turbulence-induced intensity fluctuations in FSO and intersymbol interference (ISI) in VLC due to the limited bandwidth of LEDs. In this section, in order to give insights into the effectiveness of the proposed schemes in practical OWC systems, an FSO application scenario is investigated. } 

{The specifications of the considered FSO link is given in Table \ref{table}. Note that in order to achieve $1$ Gbps data rate, a relatively large SPAD array with $4096$ pixels is employed. {In addition, a receiver aperture diameter of $10$ cm is employed to collect more received power by focusing larger received power on the detector while mitigating scintillation through aperture averaging \cite{Khalighisurvey}.} Different from the above discussion where the loss introduced by the channel is a fixed value, in FSO application the channel loss is a random variable given by  (\ref{alphaFSO}). To implement the proposed schemes in FSO, the instantaneous channel state information (CSI) $\alpha_\mathrm{FSO}$  should be known at the transmitter so that the desired transmitted photon rate vectors can be calculated through (\ref{PreDist}) or (\ref{111}). {This can be achieved by employing channel estimation schemes to estimate the channel state at the receiver and send the information back to the transmitter, for example, via a feedback path \cite{Jayaweera03}.  Alternatively, in bidirectional FSO links the channel reciprocity can be used to get the required CSI \cite{Parenti:12}. Note that in OWC systems the channel fading effects are inherently slowly fading, considering the high transmission rates the channel state remains constant over up to millions of consecutive bits \cite{Khalighisurvey}. Owing to such quasi-static channel property, channel estimation can be realized with good accuracy \cite{dabiri2017fso,Safi19}. }
In the considered FSO use case, ISI effect is assumed to be negligible due to the large bandwidth of lasers/photodetectors and the negligible dispersion effect of FSO channel \cite{Khalighisurvey}. In the simulation, for any given turbulence condition, a relatively large number of turbulence channel realizations are firstly generated. For each channel realization, the instantaneous BER performances of different systems are calculated. The corresponding average BER can then be achieved by averaging over different realizations.  Note that for the sake of simplicity, only the instantaneous BER performance calculated based on the analytical expressions is considered. }

\begin{figure}[!t]
	\centering\includegraphics[width=0.5\textwidth]{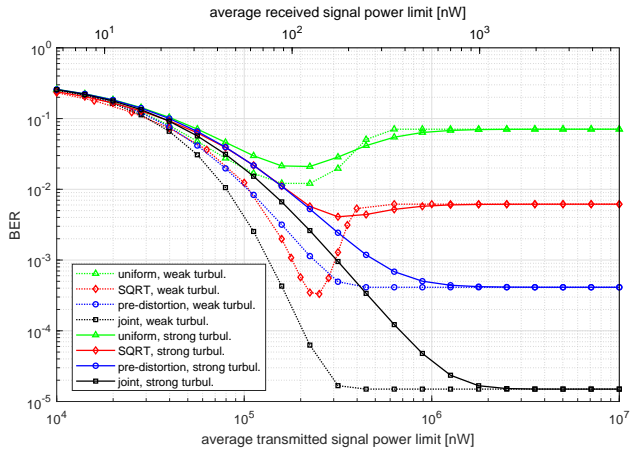}
	\caption{{{The average BER performance versus the transmitted and received signal power limits for considered systems in the application of FSO links. Two different turbulence conditions are investigated, i.e., weak turbulence with $C_n^2=10^{-15}$ $\mathrm{m}^{-2/3}$ and strong turbulence with $C_n^2=10^{-13}$ $\mathrm{m}^{-2/3}$.}}}
	\label{FSO_BER_combin_15km}
\end{figure}
{Figure \ref{FSO_BER_combin_15km} presents the simulated average BER performance versus the transmitted and received optical power for the considered systems under various turbulence conditions, i.e., $C_n^2=10^{-15}$ $\mathrm{m}^{-2/3}$ and $C_n^2=10^{-13}$ $\mathrm{m}^{-2/3}$ for weak and strong turbulence, respectively. It is presented that generally for all of the considered systems stronger turbulence condition leads to the degradation of the BER performance, as expected. For instance, for the proposed system with joint functionality, in order to achieve an average BER of $10^{-4}$, the required transmitted power under weak turbulence is only $0.2$ mW; whereas, the corresponding power under strong turbulence increases to $0.7$ mW. Hence $5.4$ dB power penalty is required in order to operate under strong turbulence. In addition, with the increase of the transmitted power the systems under weak turbulence reach the BER floor much quicker than those under strong turbulence. This is because under stronger turbulence condition, the channel loss $\alpha_\mathrm{FSO}$ experiences more significant randomness. As a result, to make sure that the average BER turns to the saturation regime where the average transmitted power is not a limiting factor for all channel realizations, much higher transmitter power is required. 
{For systems with uniform signalling and SQRT, the performance in strong turbulence could slightly outperforms that in weak turbulence when approaching the saturation. This is because at high received optical power, the SPAD operates in the nonlinear regime. The stronger turbulence induces channel gain fluctuations with larger standard deviation resulting in the SPAD receiver to operate more frequently in its linear dynamic range compared to a weaker turbulence, which leads to better performance.} {Finally, similar to the above numerical results, it is again demonstrated in Fig. \ref{FSO_BER_combin_15km}  that the the proposed systems are superior to the existing systems, which illustrates the effectiveness of the proposed systems in FSO application.}}

\section{Conclusion}\label{con}
The performance of the SPAD-based OWC systems is strongly degraded by dead-time-induced non-linear distortion and signal dependent noise. To mitigate the dead time effects and improve the communication performance, two novel systems, i.e., system with pre-distortion and system with joint pre-distortion and noise normalization, are proposed in this work which can be easily implemented without additional hardware requirements. {By comparing the proposed systems with some existing benchmark systems,} it is demonstrated that the proposed systems can significantly improve the BER performance and achievable data rate and enhance the tolerance of the background power. {In particular, it is demonstrated that when the average transmitted optical power limit is $200$ $\mu$W, with a target BER of $10^{-3}$ the proposed two systems can achieve data rates of $630$ Mbps and $900$ Mbps; whereas, the corresponding data rates of the existing systems are below $320$ Mbps.} Furthermore, the application of the proposed systems in FSO links is also investigated, which presents the effectiveness of the proposed systems in practical OWC systems.  

\appendix
{In this appendix, it is proved that for system with pre-distortion larger PAM level separation at the receiver can result in better BER performance.} 

{For the considered system, the mean value of the detected photon count is given by $\mu_{\mathrm{pre},m}=md_{\mathrm{pre}}+\xi_{\mathrm{pre}}$ with $m\in\{0, 1, \dots,M-1\}$, which is equally spaced with distance $d_{\mathrm{pre}}$. For the sake of simplicity, the subscript `pre' will be omitted in the following derivation. By substituting the mean value into (\ref{sigmam}), the variance of the signal is given by $\sigma_{m}^2=\mu_{m}-\frac{\vartheta}{N}\mu_{m}^2$. Note that it can be proved that $\mu\in[0,N/\vartheta)$, thus $\sigma_{m}^2$ is always non-negative. The optimal decoding threshold is given by (\ref{thre}) which can be further approximated as $\delta_m'=\frac{\mu_{m+1}\,\sigma_{m}+\mu_{m}\,\sigma_{m+1}}{\sigma_{m}+\sigma_{m+1}}$ \cite{elham152,gagliardi1976optical}.
By plugging this threshold into (\ref{Pebuniform}), the BER performance can be expressed as 
\begin{equation}\label{P1}
	P_\mathrm{eb}= \frac{2}{M\mathrm{log}_2(M)}\sum_{m=0}^{M-2}Q\left(\frac{\mu_{m+1}-\mu_{m}}{\sigma_{m+1}+\sigma_{m}}\right).
\end{equation}
Substituting the mean values into (\ref{P1}) gives
\begin{equation}\label{P2}
	P_\mathrm{eb}= \frac{2}{M\mathrm{log}_2(M)}\sum_{m=0}^{M-2}Q\left(\mathcal{A}(d)\right),
\end{equation}
where the function $\mathcal{A}(d)$ is given by
\begin{align}
	&\mathcal{A}(d)=\\
	&\frac{d}{\sqrt{\!(m\!+\!1)d\!+\!\xi\!-\!\frac{\vartheta}{N}\left[(m\!+\!1)d\!+\!\xi\right]^2}\!+\!\sqrt{md\!+\!\xi\!-\!\frac{\vartheta}{N}\!\left[md\!+\!\xi\right]^2}}.\nonumber
\end{align}
The first derivative of $P_\mathrm{eb}$ with respect to $d$ is given by
\begin{equation}\label{Pb1st}
	\frac{\mathrm{d}\,P_\mathrm{eb}}{\mathrm{d}\, d}=\frac{-\sqrt{2}}{\sqrt{\pi}M\mathrm{log}_2(M)}\sum_{m=0}^{M-2} e^{-\frac{1}{2}\mathcal{A}^2(d)}\, \frac{\mathrm{d}\,\mathcal{A}(d)}{\mathrm{d}\, d}.
\end{equation}
The first derivative of $\mathcal{A}(d)$ can be expressed as 
\begin{align}
&\frac{\mathrm{d}\,\mathcal{A}(d)}{\mathrm{d}\, d}=\\
&\frac{\mathcal{B}(d)}{\!\left[\!\sqrt{\!(m\!+\!1)d\!+\!\xi\!-\!\frac{\vartheta}{N}\!\left[(m\!+\!1)d\!+\!\xi\right]^2}\!\!+\!\!\sqrt{\!md\!+\!\xi\!-\!\frac{\vartheta}{N}\!\left[md\!+\!\xi\right]^2}\right]^2}\nonumber
\end{align}
where the function $\mathcal{B}(d)$  is given by 
\begin{align}
	\mathcal{B}(d)=	&\frac{md+2\xi-\frac{2\vartheta\xi}{N}(md+\xi)}{2\sqrt{md+\xi-\frac{\vartheta}{N}\left[md+\xi\right]^2}}\\
	&\qquad\quad+\frac{(m+1)d+2\xi-\frac{2\vartheta\xi}{N}\left[(m+1)d+\xi\right]}{2\sqrt{(m+1)d+\xi-\frac{\vartheta}{N}\left[(m+1)d+\xi\right]^2}}.\nonumber
\end{align}
Denoting $\mathcal{C}_g(d)=(g-\frac{2\vartheta\xi g}{N})d+2\xi-\frac{2\vartheta\xi^2}{N}$ with $g\in \{0,1,\dots, M-1\}$, $\mathcal{B}(d)$  can be rewritten as
\begin{align}\label{Bd}
	\mathcal{B}(d)=	&\frac{  \mathcal{C}_m(d) }{2\sqrt{md+\xi-\frac{\vartheta}{N}\left[md+\xi\right]^2}}\\   
	&\qquad+ \frac{ \mathcal{C}_{m+1}(d)   }{2\sqrt{(m+1)d+\xi-\frac{\vartheta}{N}\left[(m+1)d+\xi\right]^2}}.\nonumber
\end{align}
Now let's investigate the sign of function $\mathcal{C}_g(d)$. When $g=0$, we have $\mathcal{C}_0(d)=2\xi-\frac{2\vartheta\xi^2}{N}$. As $\mu\in[0,N/\vartheta)$, when $m=0$ one can get $\xi\in[0,N/\vartheta)$. Therefore,  $\mathcal{C}_0(d)\geq 0$. On the other hand, for  $g>0$, when $0<\xi\leq N/2\vartheta$ holds,  $\mathcal{C}_g(d)$ is a monotonically increasing function which a minimum value  $\mathcal{C}_g(0)=2\xi-\frac{2\vartheta\xi^2}{N}>0$. When $\xi> N/2\vartheta$ holds, $\mathcal{C}_g(d)$ is a monotonically decreasing function which a minimum value  $\mathcal{C}_g(\frac{\frac{N}{\vartheta}-\xi}{g})=\frac{N}{\vartheta}-\xi>0$. Note that as $gd+\xi$ should be less than $N/\vartheta$, the maximum value of $d$ is $\frac{\frac{N}{\vartheta}-\xi}{g}$. Finally, when $\xi=0$, $\mathcal{C}_g(d)>0$ is also satisfied. In summary, we have $\mathcal{C}_0(d)\geq 0$ and $\mathcal{C}_g(d)> 0$ when $g>0$. Thus $\mathcal{B}(d)$ given in (\ref{Bd}) is positive. As a result, $\mathcal{A}(d)$ is a monotonically increasing function with respect to $d$. According to (\ref{Pb1st}), one can conclude that the BER always decreases with the increase of $d$.  }

\bibliographystyle{IEEEtran}
\bibliography{IEEEabrv,final_manuscript}

\end{document}